\documentclass[pra, showpacs, twocolumn]{revtex4}

\usepackage[latin1]{inputenc}
\usepackage[T1]{fontenc}
\usepackage[english]{babel}

\usepackage{psfrag}
\usepackage{graphicx}
\usepackage{braket}

\usepackage[intlimits]{empheq}
\usepackage{amssymb}
\usepackage{amsthm}
\usepackage{mathtools}
\usepackage{amsfonts}

\allowdisplaybreaks[1]

\usepackage{hyperref}

\newcommand{\Axy}[2]{\hat{I}_{#1}^{\ #2}}
\newcommand{\Anm}{\hat{I}_{n}^{\ m}}
\newcommand{\Amn}{\hat{I}_{m}^{\ n}}
\newcommand{\Ann}{\hat{I}_{n}^{\ n}}

\begin{document}

\title{Superradiant phase transition in a model of three-level-lambda systems interacting with two bosonic modes}
\author{Mathias Hayn}
\author{Clive Emary}
\author{Tobias Brandes}
\affiliation{Institut f\"ur Theoretische Physik, Technische Universit\"at Berlin, 10623 Germany}
\date{\today}

\begin{abstract}
	We consider an ensemble of three-level particles in lambda-configuration interacting with two bosonic modes. The Hamiltonian has the form of a generalized Dicke-model. We show that in the thermodynamic limit this model supports a superradiant quantum phase transition. Remarkably, this can be both a first and a second order phase transition. A connection of the phase diagram to the symmetries of the Hamiltonian is also given. In addition, we show that this model can describe atoms interacting with an electromagnetic field in which the microscopic Hamiltonian includes a diamagnetic contribution. Even though the parameters of the atomic system respect the Thomas--Reiche--Kuhn sum rule, the system still shows a superradiant phase transition.
\end{abstract}

\pacs{42.50.Ct, 05.30.Rt, 32.90.+a}

\maketitle

\section{Introduction}
Superradiant phase transitions in systems of atoms interacting with an electromagnetic field were first discussed theoretically by Hepp and Lieb~\cite{Hepp73A, Hepp73}, and by Wang and Hioe~\cite{Wang73}. The superradiant phase is characterized by a macroscopic and coherent excitation of both the atoms, and the electromagnetic field modes. To date, superradiant phase transitions have been observed experimentally in artificial realizations of the seminal Dicke-model~\cite{Dicke54} in cold atoms~\cite{Baumann10}, but not for real or for artificial atoms. The question which was debated forty years ago and which is under debate today again, is whether or not superradiant phase transitions are in principle possible in these systems~\cite{Rzazewski75, Nataf10, Viehmann11}.

The system of an ensemble of atoms interacting with an electromagnetic field is often described by the Dicke-model~\cite{Dicke54}. In this model, each individual atom is described by a two-level system and the electromagnetic field by a single mode of a resonator. For the generalized Dicke-model with diamagnetic terms included (i.e. a Hopfield-like model~\cite{Hopfield58}), there exists a no-go theorem~\cite{Rzazewski75, Nataf10, Viehmann11} which precludes the transition to a superradiant phase. This no-go theorem is a consequence of sum rules and of the appearance of a diamagnetic term in the microscopic Hamiltonian which is quadratic in the transverse vector potential $\mathbf{A}$. Thus no matter how low the temperature or how strong the coupling strength is, the superradiant phase transition does not exist in such models.

The no-go theorem has also been extended to multi-level Hopfield-like-models~ \cite{Viehmann11}. However, it has been commented~\cite{Ciuti11} that these kind of no-go theorems for multi-level systems do not apply to first order superradiant phase transitions and therefore new perspectives open up.

\begin{figure}[t]
	\includegraphics[width=8.0cm]{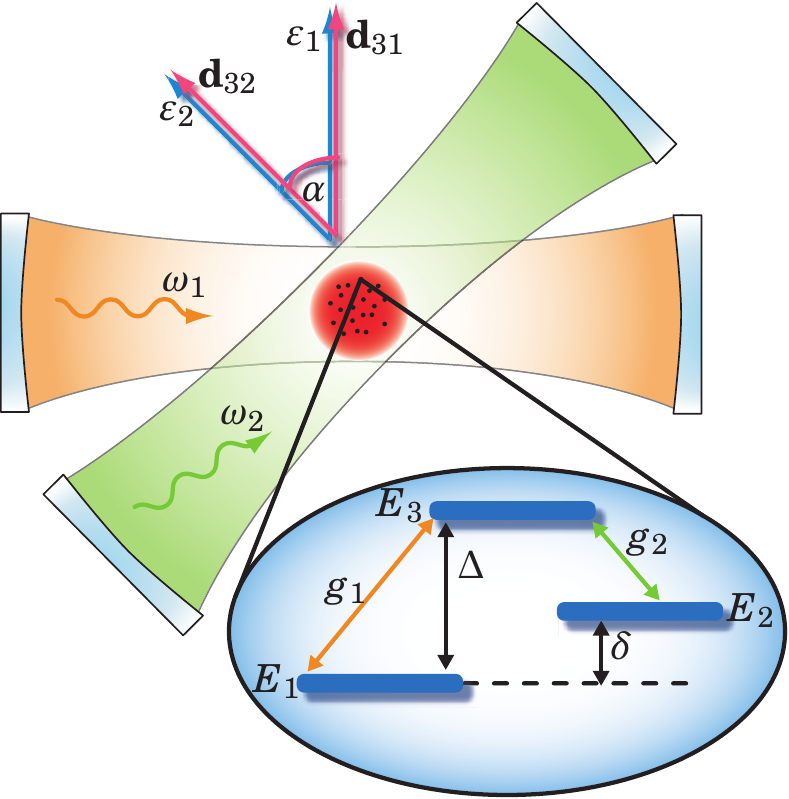}
	\caption{(Color online) Lambda-configuration of energy levels of a single particle: The two bosonic modes with energies $\omega_1$ and $\omega_2$ couple the excited state with energy $E_3$ to either of the two ground-states with energies $E_1$ and $E_2$ of the particles, respectively. The corresponding coupling strengths are given by $g_1$ and $g_2$. The geometry of the setup where the particles are atoms and the bosonic modes are modes of an electromagnetic field (cf. Sec.~\ref{sec:AModel}) is shown on the top. In our numerical analysis we assume that the polarization vectors $\mathbf{\varepsilon}_n$ are parallel to the dipole matrix elements $\mathbf{d}_{3n}$, respectively. In addition, the polarization vectors must not be orthogonal.}	\label{fig:model}
\end{figure}
In this article we extend the abstract model of our previous paper~\cite{Hayn11} by including a diamagnetic term and account for all boson mediated transitions from the ground states to the excited state. We analyze and discuss the phases and phase transitions of this extended model, and show that the system shows both a first and a second order phase transition. We also discuss the symmetries of this model and show how they are related to the quantum phases of this system. Finally, we map this model to a system consisting of atoms interacting with an electromagnetic field and show that a no-go theorem in general does \textsl{not} hold for this model and a superradiant phase transition is in fact possible. However, due to the Thomas--Reiche--Kuhn sum rule, only the first order phase transition survives. This is consistent with the results of Ref.~\cite{Baksic12}. There, three-level atoms were considered as well, and only a first order superradiant phase transition was found.

\section{The model}
We consider a system consisting of $N$ identical particles interacting with two modes of a bosonic field. The particles are described by three-level systems in lambda-configuration, i.e. two in general non-degenerate ground-states and one excited state with energies $E_1$, $E_2$, and $E_3$, and detunings $\delta = E_2 - E_1 \ge 0$ and $\Delta = E_3 - E_1 > 0$. The two bosonic modes have energies $\omega_1$ and $\omega_2$, respectively. In Fig.~\ref{fig:model}, the model is summarized graphically. The Hamiltonian is given by ($\hbar=1$)
\begin{multline}	\label{eq:Ham}
	\hat{H} = \sum_{n=1}^{3} E_n \, \Ann + \sum_{n=1}^{2} \omega_n \hat{a}_n^\dag \, \hat{a}_n \\
	+ \frac{g_1}{\sqrt{N}} \bigl( \Axy{3}{1} + \Axy{1}{3} \bigr) [ \hat{a}_1^\dag + \hat{a}_1 + \chi_1 ( \hat{a}_2^\dag + \hat{a}_2 ) ] \\
	+ \frac{g_2}{\sqrt{N}} \bigl( \Axy{3}{2} + \Axy{2}{3} \bigr) [ \hat{a}_2^\dag + \hat{a}_2 + \chi_2 ( \hat{a}_1^\dag + \hat{a}_1 ) ] \\
	+ \sum_{n=1}^2 \frac{\kappa_n^2}{\omega_n} ( \hat{a}_n^\dag + \hat{a}_n )^2 + 2 \frac{\kappa_3^2}{\sqrt{ \omega_1 \omega_2}} ( \hat{a}_1^\dag + \hat{a}_1 ) ( \hat{a}_2^\dag + \hat{a}_2 ).
\end{multline}
Here, $\Anm$ are collective atomic operators (see appendix~\ref{sec:AModel} or Ref.~\cite{Hayn11}) and $\hat{a}_n^\dag$, $\hat{a}_n$ are the creation and annihilation operators of the $n$th bosonic mode. The first two terms in Eq.~\eqref{eq:Ham} correspond to the energy of the free system of particles and a bosonic field, respectively. The third and the fourth term proportional to the coupling constants $g_1$ and $g_2$ model the interaction between the particles and the bosonic field. The dimensionless parameters $\chi_1$ ($\chi_2$) generate transitions from the first (second) ground state to the excited state induced by the second (first) bosonic mode. Eventually, the last terms scaling with $\kappa_1$, $\kappa_2$, and $\kappa_3$ represent a self interaction of the bosonic field.

Special cases of the Hamiltonian, Eq.~\eqref{eq:Ham}, include (i) the model of our previous publication~\cite{Hayn11} with $\chi_n = \kappa_n = 0$, and (ii) a model describing atoms interacting with two modes of an electromagnetic field with 
\begin{gather}
	g_n = \sqrt{N} (E_3 - E_n) \bigl| \mathbf{d}_{3n} \cdot \mathbf{\varepsilon}_n \bigr| \mathcal{A}_n, \\
	\chi_n = \frac{\alpha_{nn'}}{\alpha_{nn}} \sqrt{\frac{\omega_n}{\omega_{n'}}} \quad (n' \neq n), \\
	\kappa_1 = \kappa_2 = \kappa, \; \kappa_3 = \sqrt{\alpha} \kappa.
\end{gather}
The latter mapping and the detailed definition of the parameters is given in appendix~\ref{sec:AModel}.

By virtue of the Thomas--Reiche--Kuhn (TRK) sum rule, the couplings $g_n$ are bounded. In our model, we have two important sum rules (the derivation can be found in appendix~\ref{sec:ATRK})
\begin{equation}	\label{eq:TRK}
	g_1 \le \sqrt{ \frac{\Delta}{\omega_1} \,} \kappa \equiv g_{1,\text{TRK}} \;\; \text{and} \;\; g_2 \le \sqrt{ \frac{\Delta - \delta}{\omega_2} \,} \kappa \equiv g_{2,\text{TRK}}.
\end{equation}
Thus for given parameters, the light-matter coupling strengths $g_1$, $g_2$ for atomic systems can not exceed $g_{1,\text{TRK}}$, $g_{2,\text{TRK}}$, respectively. This bound will be crucial in the next section where we discuss phase transitions of this model.

\section{Phase transitions}

In order to obtain the phase diagram of this system, we apply the techniques presented in our previous publication~\cite{Hayn11}. There, we have used the Holstein--Primakoff-transformation~\cite{Holstein40, Klein91}. This is a non-linear transformation, which maps the collective particle operators $\Anm$ (which fulfill the algebra of generators of the unitary group) onto bosonic creation and annihilation operators. The corresponding non-linear Hamiltonian can be linearized by introducing mean-fields both for the bosonic modes, and for the creation and annihilation operators of the particles. Then, the Hamiltonian obtains the form
\begin{equation}
	\hat{H} = N \hat{h}^{(0)} + \sqrt{N \,} \hat{h}^{(1)} + \hat{h}^{(2)} + \mathcal{O}(N^{-1/2}),
\end{equation}
where the order of $N$ is explicitly given. Here, $\hat{h}^{(0)}$ corresponds to the ground-state energy, and $\hat{h}^{(2)}$ to the fluctuations around the ground-state. In the thermodynamic limit, $N \gg 1$, analysis of the ground-state energy $\hat{h}^{(0)}$ yields the relevant information for the phases, phase transitions, and phase diagram of the system. Explicitly, the ground-state energy is given by
\begin{multline}	\label{eq:GSE}
	E_0 := \hat{h}^{(0)} = \delta \Psi_2^2 + \Delta \Psi_3^2 + \biggl( \omega_1 + \frac{4 \kappa_1}{\omega_1} \biggr) \varphi_1^2 + \biggl( \omega_2 + \frac{4 \kappa_2}{\omega_2} \biggr) 
	\varphi_2^2\\ 
	+ 4 g_1 \psi_1 \Psi_3 ( \varphi_1 + \chi_1 \varphi_2 ) + 4 g_2 \Psi_2 \Psi_3 (\varphi_2 + \chi_2 \varphi_1) + 8 \frac{\kappa_3^2}{ \sqrt{\omega_1 \omega_2} } \varphi_1 \varphi_2.
\end{multline}
The real mean-fields $\varphi_s$ ($s=1,2$), $\Psi_n$ ($n=2,3$), and $\psi_1 = \sqrt{1 - \Psi_2^2 - \Psi_3^2}$ correspond to the two bosonic modes and the bosons introduced by the Holstein--Primakoff-transformation, respectively (cf. Ref.~\cite{Hayn11}). Finite values of the mean-fields give macroscopic populations of the bosonic modes or the three particle levels, i.e. if $\Psi_3$ ($\varphi_1$) is finite, then the third (first) energy level of the particles (bosonic mode) is macroscopically occupied. Particle number conservation implies $\Psi_2^2 + \Psi_3^2 \le 1$.

We eliminate the two mean-fields $\varphi_1$ and $\varphi_2$ from $E_0$ by minimizing $E_0$ with respect to these two mean-fields. Then, the ground-state energy is a function of the mean-fields $\Psi_2$ and $\Psi_3$ only: $E_0 = E_0 (\Psi_2, \Psi_3)$. One can show that the normal state with $\Psi_2 = \Psi_3 = 0$ is always a critical point of $E_0$. However, it can still be a maximum, or, if it is a minimum, it can be a local minimum only. To check whether it is a minimum or a maximum, we analyze the Hessian of $E_0 (\Psi_2 = 0, \Psi_3 = 0)$. This gives the inequality
\begin{equation}	\label{eq:Stability}
	g_1 \leq \sqrt{ \frac{\Delta  \omega_1}{4} \frac{ \Bigl( 1 + \frac{4 \kappa_1^2}{\omega_1^2} \Bigr) \Bigl( 1 + \frac{4 \kappa_2^2}{\omega_2^2}\Bigr) - \frac{16 \kappa_3^4}{\omega_1^2\omega_2^2}}{1 + \frac{4 \kappa_2^2}{\omega_2^2} + \frac{\omega_1}{\omega_2} \chi_1^2 \Bigl( 1 + \frac{4 \kappa_1^2}{\omega_1^2} \Bigr) - \frac{8 \kappa_3^2}{\omega_2^2} \chi_1 \sqrt{\frac{\omega_2}{\omega_1}}} \,} \equiv g_{1,c}.
\end{equation}
As long as this inequality is fulfilled, the normal state minimizes the ground-state energy.

\begin{figure}[t]
	\includegraphics[width=8.6cm]{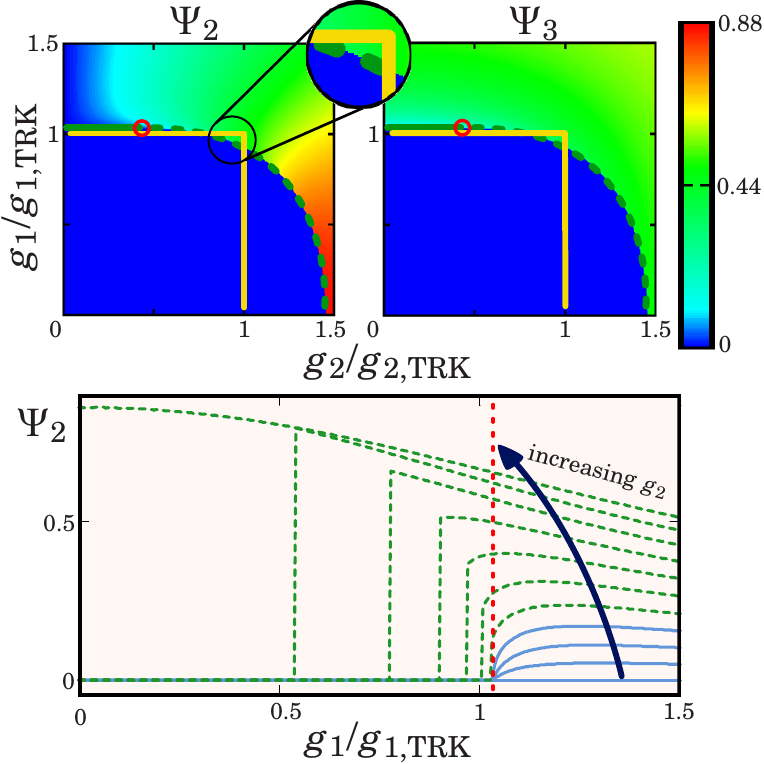}
	\caption{(Color online) \textit{Top}: Mean-fields $\Psi_2$ and $\Psi_3$ for parameters $\delta = 0.1$, $\Delta = 1$, $\omega_1 = 0.5$, $\omega_2 = 0.6$, $\alpha = \sqrt{0.5}$, $\kappa_1 = \kappa_2 = 1$, $\kappa_3 = \sqrt{\alpha}$, and $\alpha_{11} = \alpha_{22} = 1$. The green line illustrates the phase boundary and is separated into two segments: The solid (dashed) green line corresponds to a first (second) order superradiant phase transition. The point where these two segments meet, is marked with the red ring. The numerical value of $g_1$ for the boundary of the second order phase transition is given by $1.03 \, g_{1,\text{TRK}}$ which coincides with the value obtained by an analytical analysis. The blue (dark) region below the green line in both diagrams represents the normal phase ($\psi_1 = 1$, $\Psi_2 = \Psi_3 = \varphi_1 = \varphi_2 = 0$). The complementary region corresponds to the superradiant phase where all mean-fields are finite. The accessible parameter-region for atoms is indicated by the solid yellow box. The small inset clarifies that the superradiant phase is within this region. \textit{Bottom}: Mean-field $\Psi_2$ as a function of $g_1$ for different values of $g_2$, indicating the change of the order of the superradiant phase transition. The value of $g_2$ for each single line increases in direction of the arrow from $0$ to $1.5$ with an increment of $0.15$. In the range $0 \le g_2 \le 0.45$, $\Psi_2$ is continuous as a function of $g_1$ (solid blue lines), whereas for $g_2 \ge 0.6$, $\Psi_2$ is discontinuous (dashed green lines). The dashed red line marks the critical value for $g_1$ of Eq.~\eqref{eq:Stability} above which the normal phase is unstable and correspond to the red ring in the top panel. In both panels, the couplings $g_n$ are scaled with $g_{n,\text{TRK}}$. These are the largest possible couplings for the analogous atomic model given by the TRK sum rules, Eq.~\eqref{eq:TRK}. The diagrams for $\varphi_1$ and $\varphi_2$ look qualitatively the same as the diagram for $\Psi_3$. In addition, $\psi_1$ is obtained by using the relation $\psi_1 = \sqrt{1 - \Psi_2^2 - \Psi_3^2 \,}$.}	\label{fig:psi}
\end{figure}

Analyzing the Hessian of $E_0 (\Psi_2 = 0, \Psi_3 = 0)$ we can check whether or not the normal state minimizes the ground-state energy. However, we do not see if it is a global minimum of $E_0$. To see this, we need to check all critical points of $E_0$. We do this by numerical minimization of the ground-state energy $E_0$. Fig.~\ref{fig:psi} shows the order parameters $\Psi_2$ and $\Psi_3$ obtained by this numerical procedure. Regions with $\Psi_2 = 0$ and $\Psi_3 = 0$ correspond to the normal phase, whereas the superradiant phase is characterized by regions with $\Psi_2 > 0$ or $\Psi_3 > 0$. As is clear from Fig.~\ref{fig:psi}, the system supports a normal and a superradiant phase.

If two thermodynamic phases are separated by a second order phase transition the order parameter (which is given by the mean-fields) characterizing this phase transition is continuous across the phase boundary. However, numerical analysis indicates that it is not a second order phase transition along the entire phase boundary, and at a certain point the order parameters, e.g. $\Psi_2$, become discontinuous. This behavior is shown in the lower panel of Fig.~\ref{fig:psi}. Then, the phase transition is of \textsl{first} order. Consequently, the phase boundary separating the normal from the superradiant phase consists of two segments which characterize a first (solid green line in Fig.~\ref{fig:psi}) and a second (dashed green line in Fig.~\ref{fig:psi}) order phase transition, respectively. These two segments meet in a single point (red ring in Fig.~\ref{fig:psi}). The location of this point cannot be obtained by an analysis of stability of the normal phase. Likewise, we cannot derive an inequality analogous to Eq.~\ref{eq:Stability} for $g_2$.

The continuous phase transition signals the breakdown of stability of the normal phase. Therefore, the phase boundary for the second order phase transition is given by the value of $g_1$ where the inequality, Eq.~\eqref{eq:Stability}, becomes an equation, i.e. for $g_1 = g_{1,c}$. Since $g_{1,c}$ does not depend on $g_2$, the phase boundary is a straight line. With the numerical values used to generate Fig.~\ref{fig:psi}, $g_{1,c} = 1.03 \, g_{1,\text{TRK}}$ (dashed red line in Fig.~\ref{fig:psi}), where $g_{1,\text{TRK}}$ is given by Eq.~\eqref{eq:TRK}. This is in perfect agreement with our numerical findings (cf. Fig.~\ref{fig:psi}).

\begin{figure}[t]
	\includegraphics[width=8.0cm]{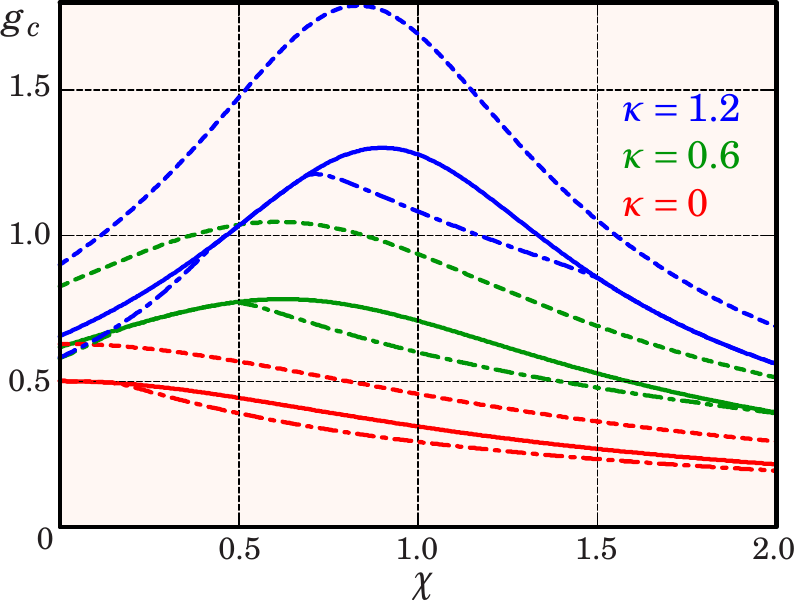}
	\caption{(Color online) Critical coupling $g_c$ at different points on the phase boundary as a function of $\chi \equiv \chi_1 = \chi_2$, and for different values of $\kappa \equiv \kappa_1 = \kappa_2 = \kappa_3$. The solid lines correspond to points on the phase boundary with $(g_1 = g_c, g_2 = 0)$, dashed lines to points with $(g_1 = 0, g_2 = g_c)$, and dashed dotted lines to points with $(g_1 = g_c, g_2 = g_c)$. The red lines (lower three) correspond to $\kappa = 0$, the green lines (three lines in center) to $\kappa = 0.6$, and the blue lines (upper three) to $\kappa = 1.2$. The remaining parameters are given by $\Delta = \omega_1 = 1$, $\delta = 0.1$, $\omega_2 = 0.9$.}	\label{fig:gcrit}
\end{figure}

It is known~\cite{Viehmann11} that the critical coupling strength increases if the diamagnetic term proportional to $\kappa$ increases. This becomes transparent, e.g., from the ground state energy, Eq.~\eqref{eq:GSE}, where the terms proportional to $\kappa_1$ ($\kappa_2$) are quadratic in $\varphi_1$ ($\varphi_2$) and thus increase the energy of the superradiant state. On the contrary, the parameters $\chi_n$ effectively give an additional contribution to the couplings $g_n$. Hence, larger parameters $\chi_n$ should lead to lower critical values of the coupling strengths. We have numerically computed the location of three points of the phase boundary for different values of $\kappa \equiv \kappa_1 = \kappa_2 = \kappa_3$ and $\chi \equiv \chi_1 = \chi_2$. This is shown in Fig.~\ref{fig:gcrit} and one clearly sees that an increase of $\kappa$ leads to an increase of $g_c$ as well, and that for large values of $\chi$ the critical coupling is lowered. We note that $g_c$ as a function of $\chi$ decreases monotonically for $\kappa = 0$ only.

Concerning the atomic system, the couplings $g_n$ are not arbitrary and can not exceed a bound given by the TRK sum rule of Eqs.~\eqref{eq:TRK}. These bounds are displayed by the solid yellow lines in Fig.~\ref{fig:psi}. As can be seen in Fig.~\ref{fig:psi}, this atomic system can in fact undergo a superradiant phase transition depending on the choice of the parameters. The numerical calculations show that this is a first order phase transition. Indeed, combining the TRK sum rule for $g_1$, Eq.~\eqref{eq:TRK}, with the stability criterion, Eq.~\eqref{eq:Stability}, for the normal state, we see that in our model the normal state is stable for all $g_1 \le g_{1,\text{TRK}}$. Hence, concerning our model, the superradiant phase transition can not be of second order. This is also in agreement with the results found for the single-mode model of Ref.~\cite{Baksic12}.

\section{Symmetries}
If one neglects the terms proportional to $\kappa_n$ ($n=1,2,3$) and $\chi_n$ ($n=1,2$), then the Hamiltonian, Eq.~\eqref{eq:Ham}, commutes with the symmetry-operators $\hat{\Pi}_n' = \exp [-i \pi (-\Ann + \hat{a}^\dag_n \, \hat{a}_n)]$, ($n=1,2$). As was shown in Ref.~\cite{Hayn11}, this gives rise to two superradiant states where a single bosonic mode is macroscopically occupied only (blue and red superradiant states), and either of the superradiant phases correspond to one broken symmetry. In contrast, including those terms, the symmetry-operators do not commute with the Hamiltonian, Eq.~\eqref{eq:Ham}, since
\begin{multline}
	\hat{\Pi}_n' \bigl( \hat{a}_m^\dag + \hat{a}_m \bigr) \bigl( \hat{a}_n^\dag + \hat{a}_n \bigr) 
	\hat{\Pi}_n^{' \dag} = \\
	\bigl( \hat{a}_m^\dag + \hat{a}_m \bigr) 
	\bigl( \hat{a}_n^\dag + \hat{a}_n \bigr) \times
	\begin{cases}
		-1&	: m \neq n, \\
		+1&	: m = n.
	\end{cases}
\end{multline}
Hence, the parities corresponding to the two branches of the lambda-system are no conserved quantities anymore, since the term proportional to $( \hat{a}_1 + \hat{a}_1^\dag ) ( \hat{a}_2 + \hat{a}_2^\dag )$ `mixes' both bosonic modes. However, the operators
\begin{equation}
\hat{\Pi}_n = \exp \bigl[ -i \pi \bigl( -\Ann + \hat{a}_1^\dag \hat{a}_1 + \hat{a}_2^\dag \hat{a}_2 \bigr) \bigr] \quad (n = 1,2)
\end{equation}
do commute with the Hamiltonian, Eq.~\eqref{eq:Ham}, and therefore this parity including both bosonic modes is conserved.

In the normal phase both parities $\hat{\Pi}_1$, $\hat{\Pi}_2$ are conserved, whereas both parities are simultaneously broken in the superradiant phase. This loss of parity symmetry has a large impact on the phase diagram: In our numerical analysis, we obtain only one superradiant phase with both bosonic modes being macroscopically excited simultaneously (cf. Fig.~\ref{fig:psi}). In addition, analytical calculations show that blue and red superradiant states are not stable. This is in contrast to the model without the terms proportional to $\kappa_n$ and $\chi_n$, where these superradiant states are the only superradiant phases and a state with both branches of the lambda-system being simultaneously superradiant is not stable~\cite{Hayn11}.

\section{Conclusion}

In this work, we have presented a generalized Dicke-model of particles with three energy-levels in lambda-configuration coupled to two bosonic modes where the microscopic Hamiltonian contains a diamagnetic term. We showed that this system exhibits a superradiant quantum phase transition in the thermodynamic limit. This phase transition can be both of first and of second order and we analytically derived the critical coupling strength for the second order phase transition. Quantitatively, the whole phase diagram was obtained using numerical methods. Compared to the model without the diamagnetic term we have studied in a previous publication \cite{Hayn11}, the phase diagram has one superradiant phase only. The loss of the second superradiant phase is directly connected to the diamagnetic term since it changes the parity symmetry of the Hamiltonian. 

In addition, we mapped this abstract model to an atomic system interacting with two photonic modes of a resonator and showed by numerical calculations that the superradiant phase transition persists. We emphasize that the microscopic Hamiltonian includes diamagnetic contributions and that this model respects the Thomas--Reiche--Kuhn sum rule which gives bounds for the coupling strengths. This is in stark contrast to Hopfield-like-models, where the combination of the diamagnetic contribution and the sum rule suppresses the superradiant phase transition. Thus, compared to the Dicke-model with diamagnetic terms included and its generalizations~\cite{Viehmann11}, no no-go-theorem exists in our model. However, we showed that the superradiant phase transition is of first order.

In experiments, our model would be realized by atoms, if they can be reduced to three-level systems in lambda-configuration. In addition, the dipole matrix element $\mathbf{d}_{31}$ ($\mathbf{d}_{32}$) must not be orthogonal to the polarization vector $\mathbf{\varepsilon}_2$ ($\mathbf{\varepsilon}_1$) of the two modes of the resonator.

In the paper of~\citet{Baksic12} a similar model is discussed. They consider atoms in a general three-level-configuration, which are coupled to one mode of a resonator. Including a diamagnetic contribution in the Hamiltonian and respecting the Thomas--Reiche--Kuhn sum rule, they find a superradiant quantum phase transition. Thus, no no-go-theorem exists in their model either. In particular for the lambda-configuration, this phase transition is always of first order. This agrees with our results.

There is one point we want to remark: Theoretically, the Dicke-model (i.e. a Hopfield-like-model without the diamagnetic term) supports a second order phase transition~\cite{Hepp73A, Wang73, Hepp73, Carmichael73, Emary03}. The no-go theorem~\cite{Rzazewski75, Nataf10, Viehmann11} applies to these continuous phase transitions only (see appendix \ref{sec:Anogo}). However, first order phase transitions could in principle still provide a superradiant phase.

\begin{acknowledgments}
	We thank A. Baksic and C. Ciuti for useful discussions. The work was supported by the Deutsche Forschungsgemeinschaft within the SFB 910 and BR 1528/8-1.
\end{acknowledgments}

\appendix
\section{Derivation of the model}
\label{sec:AModel}

In the Coulomb-gauge, the Hamiltonian of $N$ identical atoms interacting with an electromagnetic field inside a resonator is given by (cf. e.g. \cite{Grynberg89}) ($\hbar=1$)
\begin{multline}	\label{eq:Ham0}
	\hat{H} = \sum_{i=1}^{N} \Biggl( \frac{1}{2m} \sum_{j=1}^{N_e} \Bigl[ \hat{\mathbf{p}}_{ij} - q \hat{\mathbf{A}} \bigl( \hat{\mathbf{r}}_{ij} \bigr) \Bigr]^2 \\
	+ \hat{V}_c \bigl( \bigl\{ \hat{\mathbf{r}}_{i1}, \ldots, \hat{\mathbf{r}}_{iN_e} \bigr\} \bigr) \Biggr) + \sum_{s} \omega_s \,	\hat{a}_s^\dag \, \hat{a}_s.
\end{multline}
Here, $m$, $q$, $\hat{\mathbf{p}}_{ij}$ and $\hat{\mathbf{r}}_{ij}$ are the mass, the charge, the kinetic momentum and the position of the $j$th electron of $i$th atom respectively, $N_e$ is the number of electrons per atom, $\hat{\mathbf{A}}$ is the transverse vector potential, $\hat{V}_c$ is the Coulomb-potential of all electrons with respect to their respective nuclei, $\omega_s$ is the dispersion of the $s$th mode of the transverse electric field, and $\hat{a}_s^\dag$ ($\hat{a}_s$) create (annihilate) one photon in the $s$th mode. In the dipole approximation, i.e. when the wavelength of the electric field is large compared to the dimension of the system, the transverse vector potential is given by (cf. e.g. \cite{Grynberg89}) $\hat{\mathbf{A}} = \sum_s \mathcal{A}_s \mathbf{\varepsilon}_s \bigl( \hat{a}_s^\dag + \hat{a}_s \bigr)$. The quantum number $s$ contains both the polarization of the transverse electric field and the momentum $\mathbf{k}$ of the photon. Real polarization vectors $\mathbf{\varepsilon}_s$ are considered only. The parameter $\mathcal{A}_s$ is defined by
\begin{equation}
	\mathcal{A}_s = \sqrt{\frac{1}{2 \varepsilon_0 V \omega_s}},
\end{equation}
with the $\varepsilon_0$ the vacuum permittivity and $V$ the volume of the resonator.

The Hamiltonian, Eq.~\eqref{eq:Ham0}, can be written as
\begin{align}	\label{eq:Ham1}
	\hat{H} &= \sum_{i=1}^N \bigl( \hat{h}_i^{(0)} + \hat{h}_i^{(1)} \bigr) + \sum_s \omega_s \, \hat{a}_s^\dag \, \hat{a}_s, \\
	\hat{h}_i^{(0)} &= \sum_{j = 1}^{N_e} \frac{\hat{\mathbf{p}}_{ij}^2}{2m} + \hat{V}_c \bigl( \bigl\{ \hat{\mathbf{r}}_{i1}, \ldots, \hat{\mathbf{r}}_{iN_e} \bigr\} \bigr) = \sum_n E_n \ket{n}^i \hspace{-0.1cm} \bra{n}, \\
	\hat{h}_i^{(1)} &= -\frac{q}{m} \sum_{j=1}^{N_e} \hat{\mathbf{p}}_{ij} \cdot \sum_s \mathbf{\varepsilon}_s \mathcal{A}_s \bigl( \hat{a}_s^\dag + \hat{a}_s \bigr) + \\
	&+ \frac{q^2}{2m} \sum_{j=1}^{N_e} \sum_{s,s'} \mathbf{\varepsilon}_s \cdot \mathbf{\varepsilon}_{s'} \mathcal{A}_s \mathcal{A}_{s'} \bigl( \hat{a}_s^\dag + \hat{a}_s \bigr) \bigl( \hat{a}_{s'}^\dag + \hat{a}_{s'} \bigr).
\end{align}
Here, the eigensystem $\{ E_n, \ket{n}^i \}$ of the free system --- that is the kinetic energy of the $N_e$ electrons of the $i$th atom, plus the Coulomb-energy of the $N_e$ electrons with its respective nuclei --- has been introduced. The energies $E_n$ are the same for every atom. The next step is to express the kinetic momentum operator $\hat{\mathbf{p}}_{ij}$ in this basis. The identity $\hat{\mathbf{p}}_{ij} = im \bigl[ \hat{h}_i^{(0)}, \hat{\mathbf{r}}_{ij} \bigr]$ gives $\hat{\mathbf{p}}_{ij} = im \sum_{n,l} \bigl( E_n - E_l \bigr) \braket{n | \hat{\mathbf{r}}_{ij} | l} \ket{n}^j\hspace{-0.1cm}\bra{l}$.

Next, we introduce the coupling constants
\begin{gather}
	g_{n l,s} = -i \sqrt{N} \bigl( E_n - E_l \bigr) \mathcal{A}_s \mathbf{\varepsilon}_s \cdot \mathbf{d}_{n l}, \\
	\text{with } \mathbf{d}_{n l} = q \sum_{j=1}^{N_e} \braket{n | \hat{\mathbf{r}}_{ij} | l},
\end{gather}
where the dipole matrix elements $\mathbf{d}_{n l}$ are identical for all atoms. Then, the Hamiltonian, Eq.~\eqref{eq:Ham1}, assumes the form
\begin{multline}	\label{eq:Ham2}
	\hat{H} = \sum_n E_n \Ann + \sum_s \omega_s \, \hat{a}_s^\dag \, \hat{a}_s \\
	+ \sum_s \sum_{n > l} \biggl( \frac{g_{n l,s}}{\sqrt{N}} \Axy{n}{l} + \frac{g_{n l,s}^*}{\sqrt{N}} \Axy{l}{n} \biggr) \bigl( \hat{a}_s^\dag + \hat{a}_s \bigr) \\
	 + \sum_{s,s'} \frac{\kappa^2}{\sqrt{\omega_s \omega_{s'}}} \, \mathbf{\varepsilon}_s \cdot \mathbf{\varepsilon}_{s'} \bigl( \hat{a}_s^\dag + \hat{a}_s \bigr) \bigl( \hat{a}_{s'}^\dag + \hat{a}_{s'} \bigr),
\end{multline}
with collective operators $\Axy{n}{l} = \sum_{i=1}^N \ket{n}^i\hspace{-0.1cm}\bra{l}$, and the parameter $\kappa$ is given by
\begin{equation}	\label{eq:DefKappa}
	\kappa = \sqrt{\frac{q^2 N N_e \omega_n}{2m}} \mathcal{A}_n.
\end{equation}

Finally, we restrict the Hamiltonian, Eq.~\eqref{eq:Ham2}, to a three-level lambda-system with two modes of a resonator, i.e. $n=1,2,3$ and $s=1,2$. We assume that the dipole matrix element $\mathbf{d}_{12} = 0$. This may be due to symmetry. Eventually, introducing real coupling constants
\begin{equation}	\label{eq:DefCouplings}
	g_n = \sqrt{N} (E_3 - E_n) \bigl| \mathbf{d}_{3n} \cdot \mathbf{\varepsilon}_n \bigr| \mathcal{A}_n, \quad (n =1,2)
\end{equation}
the Hamiltonian of Eq.~\eqref{eq:Ham} with $\chi_n = \frac{\alpha_{n l}}{\alpha_{nn}} \sqrt{\frac{\omega_n}{\omega_l}}$ ($n \neq l = 1,2$), $\alpha_{nl} = \bigl| \mathbf{d}_{3n} \cdot \mathbf{\varepsilon}_l \bigr| \bigl/ | \mathbf{d}_{3n} |$, $\kappa_1 = \kappa_2 = \kappa$, $\kappa_3 = \kappa \sqrt{\alpha}$, and $\alpha = \mathbf{\varepsilon}_1 \cdot \mathbf{\varepsilon}_2$ is obtained.

\section{Derivation of the Thomas--Reiche--Kuhn sum rules}
\label{sec:ATRK}

For every Hamiltonian $\hat{H}$ with spectrum $E_n$ and eigenbasis $\ket{n}$ the identity
\begin{equation}	\label{eqA:SumRule}
	\sum_n \bigl(E_n - E_l \bigr) \braket{l | \hat{O} | n} \braket{n | \hat{O} | l} = \frac{1}{2} \braket{l | \bigl[ \hat{O}, [\hat{H}, \hat{O}] \bigr] | l}
\end{equation}
is fulfilled. Here, the sum runs over all quantum numbers $n$. This equation is valid for any Operator $\hat{O}$ and every quantum number $l$ of the Hamiltonian $\hat{H}$.

A special sum rule for atomic systems with a Hamiltonian of the form
\begin{equation}
	\hat{H} = \sum_{j=1}^{N_e} \frac{\hat{\mathbf{p}}_{ij}^2}{2m} + \hat{h} \bigl( \bigl\{ \hat{\mathbf{r}}_{i1}, \ldots, \hat{\mathbf{r}}_{iN_e} \bigr\} \bigr)
\end{equation}
and operators
\begin{equation}
	\hat{O} = \mathbf{\varepsilon} \cdot \hat{\mathbf{d}}, \quad \hat{\mathbf{d}} = q \sum_{j=1}^{N_e} \hat{\mathbf{r}}_{ij}, \quad |\mathbf{\varepsilon}| = 1
\end{equation}
is given by
\begin{equation}	\label{eqA:TRK}
	\sum_n \bigl( E_n - E_l \bigr) \, \bigl| \mathbf{\varepsilon} \cdot \mathbf{d}_{l n} \bigr|^2 = \frac{q^2 N_e}{2m}, \; \mathbf{d}_{n l} = \braket{n | \hat{\mathbf{d}} | l}.
\end{equation}
This holds for all quantum numbers $l$. This kind of sum rule for the dipole matrix element $\mathbf{d}_{n l}$ is called Thomas--Reiche--Kuhn-sum rule~\cite{Bethe86}.

Since in our model the infinite dimensional Hilbert-space of a single atom is restricted to three energy levels only, and $\mathbf{d}_{12} = 0$ is assumed, the left side of Eq.~\eqref{eqA:TRK} is bounded, i.e.
\begin{equation}
	\sum_{n=1}^{3} \bigl( E_n - E_l \bigr) \, \bigl| \mathbf{\varepsilon} \cdot \mathbf{d}_{n l} \bigr|^2 \le \frac{q^2 N_e}{2m}, \; l = 1,2.
\end{equation}
Eventually, using the definitions for $g_n$ (Eq.~\eqref{eq:DefCouplings}) and for $\kappa$ (Eq.~\eqref{eq:DefKappa}), and choosing $l = 1, 2$, we obtain the two inequalities of the Eqs.~\eqref{eq:TRK}.

\section{No-go theorem for second order superradiant phase transitions}
\label{sec:Anogo}

We consider the most general model of an ensemble of $N$ multi-level atoms interacting with a single mode of a resonator. The Hamiltonian has the form
\begin{multline}
	\hat{H} = \sum_{n=1}^\nu E_n \Ann + \omega \hat{a}^\dag \, \hat{a} + \frac{\kappa^2}{\omega} (\hat{a}^\dag + \hat{a})^2 \\
	+ \sum_{n,m=1}^\nu \frac{1}{2} \frac{g_{n,m}}{\sqrt{N \,}} \bigl( \Anm + \Amn \bigr) (\hat{a}^\dag + \hat{a}).
\end{multline}
Each single atom has $\nu$ energy levels characterized by the non-degenerate energies $E_1 < E_2 < \ldots < E_\nu$. Populations of and transitions among the atomic energy levels are described by the collective operators $\Ann$ and $\Anm$, respectively. The energy of the mode of the resonator is given by $\omega > 0$; the bosonic operator $\hat{a}^\dag$ ($\hat{a}$) create (annihilate) a corresponding photon. The coupling of the transverse vector potential with the atoms is approximated by the dipole coupling, with coupling strength $g_{n,m} = g_{m,n}$, $g_{n,n} = 0$. The diamagnetic contribution is parameterized via the parameter $\kappa$. In deriving this Hamiltonian, we have neglected atom-atom and Coulomb-interaction among different atoms.

The TRK sum rules (cf. Eq.~\eqref{eqA:TRK}) of this model can be written in the form
\begin{equation}	\label{eqC:TRK}
	\sum_{\substack{n=1 \\ n \neq m}}^\nu \frac{g_{n,m}^2}{E_{n,m}} \le \frac{\kappa^2}{\omega},
\end{equation}
with $E_{n,m} = E_n - E_{m}$ and $m = 1, \ldots, \nu$. Note that in terms of oscillator strengths $f_{n,m} = \frac{g_{n,m}^2 \omega}{E_{m,n} \kappa^2}$, the TRK sum rule obtains the memorable form $\sum_{n} f_{n,m} = 1$.

Applying a Holstein--Primakoff-transformation to the collective operators $\Anm$ and introducing mean fields both for the atomic [$\Psi_n$, ($n=2,\ldots,\nu$)] and for the photonic ($\varphi$) degrees of freedom, we obtain the ground-state energy per atom in the thermodynamic limit
\begin{multline}	\label{eqC:GSE1}
	E_{GS} = E_1 + \sum_{n=2}^\nu E_{n,1} \Psi_n^2 + \left( \omega + 4 \frac{\kappa^2}{\omega} \right) \varphi^2 \\
	+ 2 \varphi \sum_{n=2}^\nu \left( 2g_{n,1} \sqrt{1 - \sum_{m=2}^\nu \Psi_m^2} + \sum_{m=2}^\nu g_{n,m} \Psi_m \right) \Psi_n.
\end{multline}
Since we are only interested whether or not the normal state with $\Psi_n = \varphi = 0$ minimizes the ground-state energy, Eq.~\eqref{eqC:GSE1}, we expand the ground-state energy to second in order in $\Psi_n$ and $\varphi$. This yields
\begin{multline}	\label{eqC:GSE2}
	E_{GS} = E_1 + \sum_{n=2}^\nu E_{n,1} \Psi_n^2 + \left( \omega + 4 \frac{\kappa^2}{\omega} \right) \varphi^2 \\
	+ 4 \sum_{n=2}^\nu g_{n,1} \Psi_n \varphi + \mathcal{O} (\Psi_n^3).
\end{multline}
Because the ground-state energy has no terms linear in $\Psi_n$ or $\varphi$, the normal state represents a critical point. In order to specify the type of the critical point, we calculate the Hessian of the ground-state energy, Eq.~\eqref{eqC:GSE2}. We obtain
\begin{align}
	\frac{\partial^2 E_{GS}}{\partial \varphi^2} &= 2 \left( \omega + 4 \frac{\kappa^2}{\omega} \right) + \mathcal{O} (\Psi_n^3), \\
	\frac{\partial^2 E_{GS}}{\partial \Psi_m \, \partial \Psi_n} &= 2 E_{n,1} \delta_{n,m} + \mathcal{O} (\Psi_n), \text{ and} \\
	\frac{\partial^2 E_{GS}}{\partial \varphi \, \partial \Psi_n} &= 4 g_{n,1} + \mathcal{O} (\Psi_n^2)
\end{align}
with $1 < n,m \le \nu$. Hence, the Hessian for the normal state is given by
\begin{multline}
\mathbf{H} \bigl( E_{GS}, \Psi_n = 0, \varphi = 0 \bigr) = \\
2
\begin{pmatrix}
	\omega + 4 \frac{\kappa^2}{\omega} & 2 g_{2,1} & 2 g_{3,1} & \dots & 2 g_{\nu-1,1} & 2 g_{\nu,1} \\
	2 g_{2,1} & E_{2,1} & 0 & \dots & 0 & 0 \\
	2 g_{3,1} & 0 & E_{3,1} & \dots & 0 & 0 \\
	\vdots & \vdots & \vdots & \ddots & \vdots & \vdots \\
	2 g_{\nu-1,1} & 0 & 0 & \dots & E_{\nu-1, 1} & 0 \\
	2 g_{\nu,1} & 0 & 0 & \dots & 0 & E_{\nu, 1}
\end{pmatrix}.
\end{multline}
Now we will proof that $\mathbf{H}$ is in fact positive definite which means that the critical point is a local minimum. Therefore we compute the principal minors of $\mathbf{H}$. The $k$th principal minor of a $n \times n$ matrix $\mathbf{M}$ is the determinant of the matrix where the first $n - k$ rows and columns of $\mathbf{M}$ are deleted. If all principal minors of a symmetric matrix are positive, then the matrix is positive definite. In the following we will show that all principal minors of $\mathbf{H}$ are positive.

One sees readily that the first $\nu-1$ principal minors are positive since $E_{n,1}$ are positive by definition. Hence, we need to compute the $\nu$th principal minor which is the determinant of $\mathbf{H}$ itself. This is done by reducing $\mathbf{H}$ to a triangular matrix using elementary row operations. Eventually, this gives
\begin{multline}
\frac{1}{2} \mathrm{Det} \bigl[ \mathbf{H} \bigl( E_{GS}, \Psi_n = 0, \varphi = 0 \bigr) \bigr] = \\
\begin{vmatrix}
	\mathrm{X} & 0 & 0 & \dots & 0 & 0 \\
	2 g_{2,1} & E_{2,1} & 0 & \dots & 0 & 0 \\
	2 g_{3,1} & 0 & E_{3,1} & \dots & 0 & 0 \\
	\vdots & \vdots & \vdots & \ddots & \vdots & \vdots \\
	2 g_{\nu-1,1} & 0 & 0 & \dots & E_{\nu-1,1} & 0 \\
	2 g_{\nu,1} & 0 & 0 & \dots & 0 & E_{\nu,1},
\end{vmatrix}
\end{multline}	
where $\mathrm{X} = \omega + 4 \frac{\kappa^2}{\omega} - 4 \sum_{n=2}^\nu \frac{g_{n,1}^2}{E_{n,1}}$. Applying the TRK sum rule, Eq.~\eqref{eqC:TRK}, with $m = 1$, we obtain
\begin{equation}
	\mathrm{X} \ge \omega
\end{equation}
which is always positive. Since the determinant of a triangular matrix is the product of its diagonal elements, the determinant of the Hessian for the normal state is positive. Hence we have shown that all principal minors are positive. Consequently, the Hessian is positive definite. This means that the normal state minimizes the ground-state energy irrespective of the parameters of the Hamiltonian and no additional states, i.e. superradiant states, can evolve from the normal state continuously. Thus, no continuous, i.e. second order, phase transitions are possible. However, this argument does not apply to first order phase transitions, and we can not say whether or not first order phase transitions can occur. In addition, a second order superradiant phase transition originating from a superradiant phase which results from a first order superradiant phase transition is also not covered by our analysis. We also note that a superradiant phase where a single one-particle energy level is macroscopically occupied only can be excluded by a similar argument as presented above.

In summary, we have considered the most general model of multi-level atoms interacting with a single mode of a resonator which obey the Thomas--Reiche--Kuhn sum rule. We have shown that the normal state, the state where all atoms occupy their respective ground state and in the resonator no photon is excited, does always minimize the ground state energy. Hence the normal phase is stable irrespective of the parameters of the system and no second order phase transitions to superradiant phases are possible.

\end{document}